\RequirePackage{fix-cm}
\documentclass[twocolumn,epjc3]{svjour3}  
\smartqed  
\RequirePackage{graphicx}

\usepackage{gensymb}                   
\usepackage{bbold}
\usepackage{graphicx}
\usepackage{amsmath}
\usepackage{textcomp}
\usepackage{dcolumn}
\usepackage{bm}
\usepackage{amssymb}
\usepackage{amsfonts}
\usepackage{lineno}
\usepackage{url}
\usepackage{array}
\usepackage{verbatim}
\usepackage{subfigure}
\usepackage{listings}
\usepackage{color}
\usepackage{lscape}
\usepackage{setspace}
\usepackage{footmisc}
\usepackage [english]{babel}
\usepackage [autostyle, english = american]{csquotes}

\MakeOuterQuote{"}
\modulolinenumbers[5]

\journalname{Eur. Phys. J. Appl. Phys.}
\begin{document}

\title{Statistical Uncertainty in Quantitative Neutron Radiography}


\author{F.~M.~Piegsa\thanksref{e1,addr1}
        \and
        A.~P.~Kaestner\thanksref{e2,addr2} 
				\and
				A.~Antognini\thanksref{addr1,addr2}
				\and
				A.~Eggenberger\thanksref{addr1}
				\and
				K.~Kirch\thanksref{addr1,addr2}
				\and
				G.~Wichmann\thanksref{addr1}
}

\thankstext{e1}{e-mail: florian.piegsa@phys.ethz.ch}
\thankstext{e2}{e-mail: anders.kaestner@psi.ch}


\institute{ETH Z{\"u}rich, Institute for Particle Physics, CH-8093 Z{\"u}rich, Switzerland \label{addr1}
           \and
           Paul Scherrer Institute, CH-5232 Villigen, Switzerland \label{addr2}
}

\date{Received: date / Accepted: date}

\maketitle

\begin{abstract}
We demonstrate a novel procedure to calibrate neutron detection systems commonly used in standard neutron radiography. This calibration allows determining the uncertainties due to Poisson-like neutron counting statistics for each individual pixel of a radiographic image. 
The obtained statistical errors are necessary in order to perform a correct quantitative analysis. This fast and convenient method is applied to data measured at the cold neutron radiography facility ICON at the Paul Scherrer Institute. Moreover, from the results the effective neutron flux at the beam line is determined.

\end{abstract}

\section{Introduction}

Neutron radiography and tomography represent powerful non-destructive imaging techniques \cite{Anderson2009}. They are applied in a variety of different fields of science ranging from physics investigations to engineering, cultural heritage, biology etc. A recent comprehensive overview covering the entire spectrum of research applications is given in the proceedings of the 10$^{\text{th}}$ World Conference on Neutron Radiography \cite{WCNR10}. A list of neutron radiography facilities around the world in operation using spallation and reactor sources can be found in \cite{ISNR-web,Lehmann201510}.\\
Until now, most neutron radiography experiments aim at quantifying information based on the transmission through the samples or shapes present in the images. For this kind of quantification the absolute number of neutrons and the corresponding statistical uncertainty are less relevant. 
However, real quantitative data assessment involving dedicated and extensive analysis procedures, e.g.\ curve-fitting or interpretation of scattering data, is becoming more and more important. Examples of such applications encompass for instance polarized neutron radiography \cite{Kardjilov2008,Strobl2009,Strobl20112415,Treimer2014,Tremsin2015,Piegsa2008a,Piegsa2009a,Piegsa2011a} and various novel techniques with an explicit overlap of neutron scattering and radiography, e.g.\ grating interferometry or multiple small angle neutron scattering etc. \cite{Grunzweig2008-1,Grunzweig2013-1,Manke2010,Reimann2015,Strobl2008,Strobl2015,Grunzweig2007,Pierret201239,Betz2016}. 
For the quantitative analysis using these methods it is inevitable to assign proper statistical errors to the measured data values for each individual pixel.\footnote{The problem appeared already in one of our earlier publications investigating a $^3$He gas density gradient at low temperatures \cite{Wichmann2016}. Here, we present a dedicated and systematic study of the developed calibration procedure.}
This is not only essential in order to perform a valid regression analysis with weighted data points, but also to optimize the achievable signal-to-noise ratio and thus ultimately the image quality. The latter is especially important in the realm of low-dose neutron radiography, i.e.\ dynamic, fast, and/or high-resolution imaging \cite{kaestner2015_SED,Trtik2015169,Vontobel2005148}. Furthermore, this knowledge allows then for a detailed off-line evaluation of the feasibility of planned experiments and investigations. \\
The challenge lies in transferring the measured intensity in arbitrary units of the two-dimensional detector, usually given in grayscale values of a CCD-camera, into actual neutron counts with a corresponding statistical error.
In the present article, we describe a procedure which allows for a calibration of a standard neutron imaging system without any additional detection equipment, i.e.\ a $^3$He gas detector or gold foil activation analysis.
The method is demonstrated at a cold neutron source, however, the results can be easily adopted to a radiography facility with a different neutron energy spectrum.

\section{Description of the Method}

\begin{figure}
	\centering
		\includegraphics[width=0.45\textwidth]{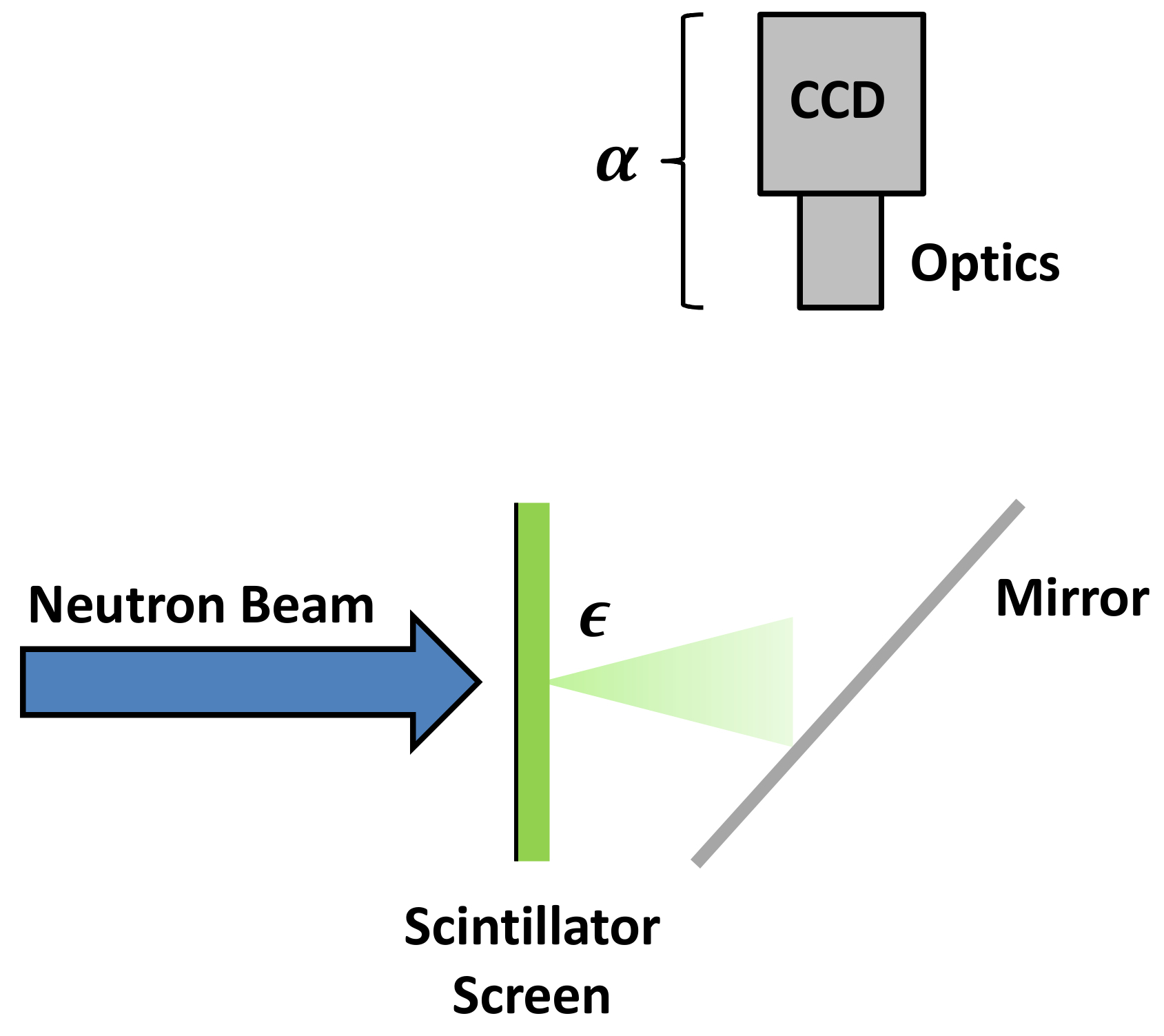}
	\caption{Scheme of a standard neutron radiography detection system. Neutrons hitting the scintillation screen are producing scintillation light with an efficiency $\epsilon$. The light is reflected from a mirror to a CCD-camera. The process of light detection and conversion to grayscale intensities with its according efficiencies is described by the scaling factor $\alpha$. }
	\label{fig:Fig01_setup}
\end{figure}
For our method, we consider a neutron radiography detection system consisting of a neutron-absorbing scintillator screen which is recorded by means of a low-noise CCD-camera 
(compare Fig.\ \ref{fig:Fig01_setup}). However, any other detection system with similar properties can be employed equivalently, e.g.\ a neutron imaging plate \cite{Kobayashi19991,Tazaki199920}.
The analysis is resting upon four plausible assumptions: 
(a)~The neutron counting statistics is Poisson distributed and for a large number of events the shape of this distribution can be approximated by a Gaussian. 
(b)~The measured intensity is caused only by the neutrons, hence, we do not take into account any potential background due to $\gamma$-radiation. 
(c)~As will be presented below, using modern CCD-cameras with a photo-sensitive chip cooled to low temperatures the so-called dark current and its corresponding noise are completely negligible. The same is true for the read-out noise of the CCD.
(d)~The grayscale intensity $I$ is linearly proportional to the number of absorbed neutrons. The last assumption translates into the following equation:
\begin{equation}
	I = \alpha \epsilon N
	\label{eq:intensity}
\end{equation}
where $\alpha$ is a scaling factor of the detection/camera system describing the conversion of light emitted by the scintillation screen to photo-electrons in the CCD-chip.
Moreover, $\epsilon$ is the detection efficiency of the neutrons in the scintillator and $N$ is the number of neutrons hitting the screen, respectively.
Thus, the number of detected neutrons is $\epsilon N$ and the corresponding statistical uncertainty due to Poisson statistics is given by $\sqrt{\epsilon N}$ \cite{Buzug2008}. As a consequence, we find for the uncertainty of the intensity:
\begin{equation}
	\delta I = \alpha \sqrt{\epsilon N}
	\label{eq:intensity-error1}
\end{equation}
Alternatively, by combining Eq.\ (\ref{eq:intensity}) and (\ref{eq:intensity-error1}) we obtain:
\begin{equation}
  \delta I = \sqrt{\alpha I} 
\label{eq:intensity-error2}
\end{equation}
While these three equations are valid for each individual pixel of an image, we further assume that $\alpha$ and $\epsilon$ do not depend on the position, i.e.\ $\epsilon$ has the same value everywhere on the scintillator and $\alpha$ is constant over the entire CCD-chip. This is another reasonable presumption which can be experimentally confirmed by exposing different areas of the scintillator with the same neutron beam.
From Eq.\ (\ref{eq:intensity-error2}) it becomes obvious that in order to calculate the statistical error $\delta I$  for each pixel it is necessary to determine the scaling factor $\alpha$. Furthermore, if $\alpha$ is known one can also directly infer on the number of detected neutrons via the measured intensity using 
Eq.\ (\ref{eq:intensity}). \\
To cancel out intensity variations or inhomogeneities of the neutron beam over the field-of-view of the detection system, one usually performs a radiometric normalization of the images with respect to a so-called open beam image with much higher statistics, i.e.\ an average of many images taken at long exposure times.  Here, this will be described by a normalization intensity $I_0=\alpha \epsilon N_0$. Hence, the relative intensity and relative uncertainty of the intensity are given by:
\begin{equation}
  \frac{I}{I_0} = \frac{N}{N_0}  \text{ }\text{ }\text{ }\text{ }\text{ }\text{ and }\text{ }\text{ }\text{ }\text{ }\text{ }   \frac{\delta I}{I_0} = \frac{\sqrt{\alpha}}{\sqrt{I_0}} \cdot 	\sqrt{\frac{I}{I_0}}
\label{eq:relativeintensity}
\end{equation}
The gist of the new procedure is to determine $\alpha$ by performing a set of measurements from which one extracts $\delta I/I_0$ as a function of $I/I_0$. This is done by taking images of the neutron beam for various exposure times causing different accumulated intensities. Since the exposure with neutrons is a statistical process the intensity values of a large number of pixels will scatter around their mean value with a Gaussian distribution.\footnote{For very low intensities or count rates this might be also described by the related Rician distribution. This has been reported in an investigation concerning the signal-to-noise ratio in x-ray dark-field imaging \cite{chabior2011_InterferometerNoise}.}  
The width of this distribution arises from a contribution caused by the neutron Poisson statistics and the beam inhomogeneity over the considered area of pixels. However, the latter is canceled out by normalizing with the open beam image. Thus, the residual width then directly corresponds to the relative uncertainty $\delta I/I_0$.

\section{Experiment and Data Analysis}

\begin{figure}
	\centering
		\includegraphics[width=0.48\textwidth]{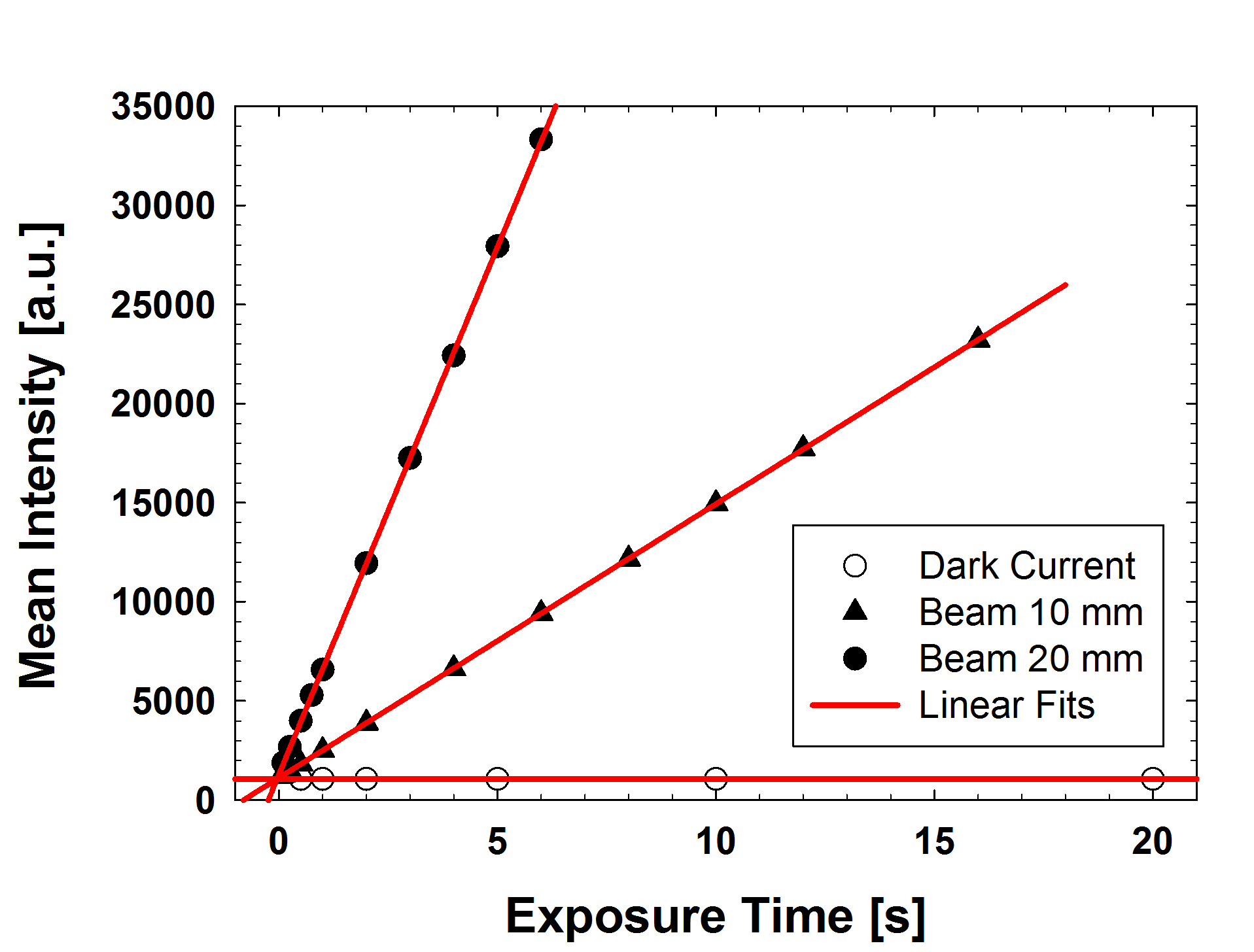}
	\caption{Mean intensity per pixel averaged over an area of $200 \times 200$~pixels in arbitrary units (grayscale values) taken at a SINQ proton beam current of 1.5~mA. From the slope of the linear fits (lines) one can extract the mean accumulated intensity per exposure time. }
	\label{fig:Fig01_Intensity}
\end{figure}
In the following, example measurements and the corresponding data analysis are presented employing the aforementioned method. The data has been collected at the ICON imaging beam line at the continuous spallation neutron source SINQ at the Paul Scherrer Institute in Switzerland \cite{Kaestner2011387}.
The radiographic images were taken in the classic pinhole geometry with an aperture-to-detector distance of approximately 6.5~m. A scintillator screen made from $^6$Li/ZnS:Cu with a thickness of 100~\textmu m was used \cite{Spowart196935}. The scintillation light was recorded with an ANDOR iKon-L CCD-camera with a 16~bit dynamic range and a $2048 \times 2048$~pixels photo-sensitive chip cooled to $-75$\degree C \cite{ANDOR-web}. With the camera, a Nikkor 50~mm f/1.4 lens was employed. The measured effective pixel size of the obtained neutron images was approximately $A_{\text{eff}} = 0.1 \times 0.1$~mm$^2$. \\
Firstly, a collection of so-called dark current images, i.e.\ with the neutron beam off, were taken for different exposure times of up to 60~s. This data is used to investigate the intrinsic offset/bias and noise of the CCD-chip. To ensure that the measured offset is not increased by a potential afterglow of the scintillator the detector has not been exposed to neutrons for at least several hours prior to these measurements.\\
Secondly, images were taken at a SINQ proton beam current of $I_{\text{SINQ}} = 1.5$~mA.
In order to perform a systematic investigation, data was acquired for two different neutron beam aperture (pinhole) settings, namely a diameter $d$ of 10~mm and 20~mm, respectively. In the analysis, only an area of $200 \times 200$ pixels, i.e.\ $2 \times 2$~cm$^2$, in the center of the images is considered which experienced a very homogeneous neutron beam exposure. 
In Fig.\ \ref{fig:Fig01_Intensity} the mean recorded intensity (in grayscale values) per pixel averaged over this area is plotted as a function of the exposure time. 
The offset obtained from the dark current measurements exhibits a constant value of $1058 \pm 1$, even for very long exposure times.\footnote{A linear fit to the dark current data yields actually a very small intensity increase of $\dot{I}_{\text{DC}} = (0.02 \pm 0.01)$~s$^{-1}$.} 
The linear fits to the data points with the neutron beam shutter open yield a mean accumulated intensity per exposure time, i.e.\ the slope of the linear fit, of 
$\dot{I}_{\text{10}} = (1381 \pm 1)$~s$^{-1}$ for an aperture diameter of 10~mm and 
$\dot{I}_{\text{20}} = (5321 \pm 11)$~s$^{-1}$ for an aperture diameter of 20~mm, respectively.\footnote{All given uncertainties correspond to the standard errors of a linear regression without weighting of the data points.} As expected, the intensity scales roughly with the cross section of the neutron beam aperture, i.e.\ with its diameter $d$ squared.
\begin{figure}
	\centering
		\subfigure[]{\includegraphics[width=0.45\textwidth]{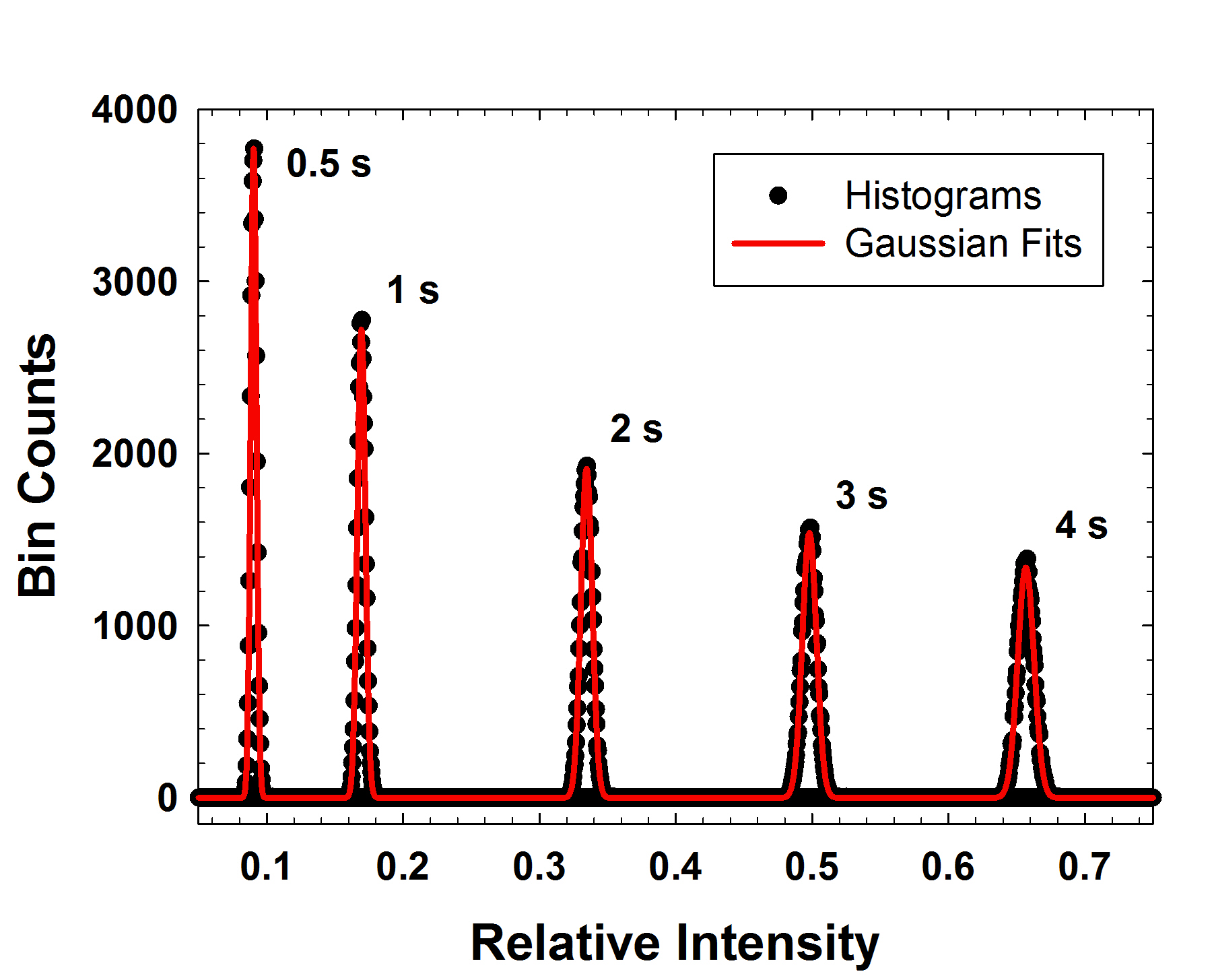} } \\
		\subfigure[]{\includegraphics[width=0.45\textwidth]{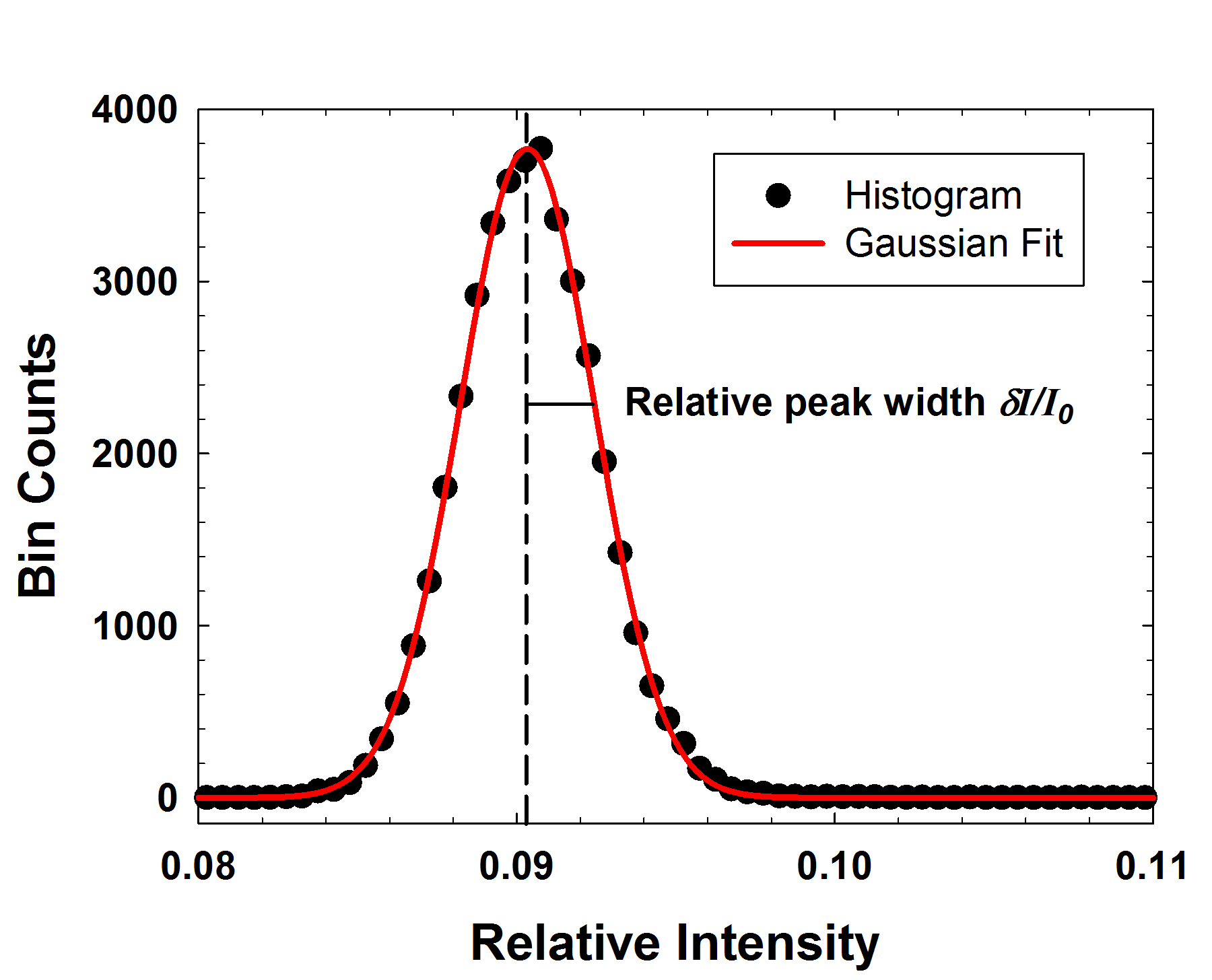} }
	\caption{Top: Histograms of the relative intensity $I / I_0$ of the central $200 \times 200$~pixel area for different exposure times (0.5~s, 1~s, 2~s, 3~s, and 4~s) at an aperture diameter of 20~mm. The individual images have been normalized to the average of a set of 10 images with an exposure time $t_{\text{N}} = 6$~s. The bin width of the plot is $5 \times 10^{-4}$. The lines represent Gaussian fits to the peaks in the data. Bottom: The same plot zoomed in onto the histogram peak corresponding to the image exposed for 0.5~s. Here, the Gaussian fit yields a maximum of the relative intensity at 
	$I / I_0 \approx 0.09$ (dashed vertical line) and a relative peak width (standard deviation/short horizontal line) of $\delta I / I_0 = 2.1 \times 10^{-3}$. }
	\label{fig:Fig02_Histograms20mm}
\end{figure}\\
In addition, two sets of 10 images each were recorded with long exposure times $t_{\text{N}}$ of 10~s for the 10~mm and 6~s for the 20~mm beam aperture, respectively.\footnote{Note, the maximum exposure time is limited by the saturation of the CCD-chip.}
The pixel-wise averages of these sets serve as references for the normalization of the aforementioned images exposed for shorter times. Moreover, the constant offset of 1058 is subtracted from each evaluated image to compensate for the CCD-bias. 
These offset corrected and normalized versions of the initial images 
are further investigated by creating histograms of their relative intensity $I / I_0$, again only considering the central $200 \times 200$ pixel area. 
Thus, each histogram represents the spread of the pixel intensities of the 40000 pixels for a certain exposure time. 
As an example, the results for some of the images taken with the 20~mm beam aperture are presented in Fig.\ \ref{fig:Fig02_Histograms20mm}.
The individual histograms exhibit a peak centered around the exposure time ratio, e.g.\ the image exposed for 3~s has a mean relative intensity of approximately 0.5 with respect to the normalization with 
$t_{\text{N}} = 6$~s. The width of the peaks can be determined by applying a Gaussian fit to the data. As expected, due to the Poisson nature of the neutron statistic with increasing relative intensity the width of these peaks become larger. Note, that the aforementioned intrinsic dark current noise and read-out noise of the CCD-chip do not cause an additional broadening of the peaks, since their contribution is in the range of $10^{-4}$ and, thus, at least one-to-two orders of magnitude smaller than the measured contribution due to the neutron statistics.
\begin{figure}
	\centering
		\includegraphics[width=0.45\textwidth]{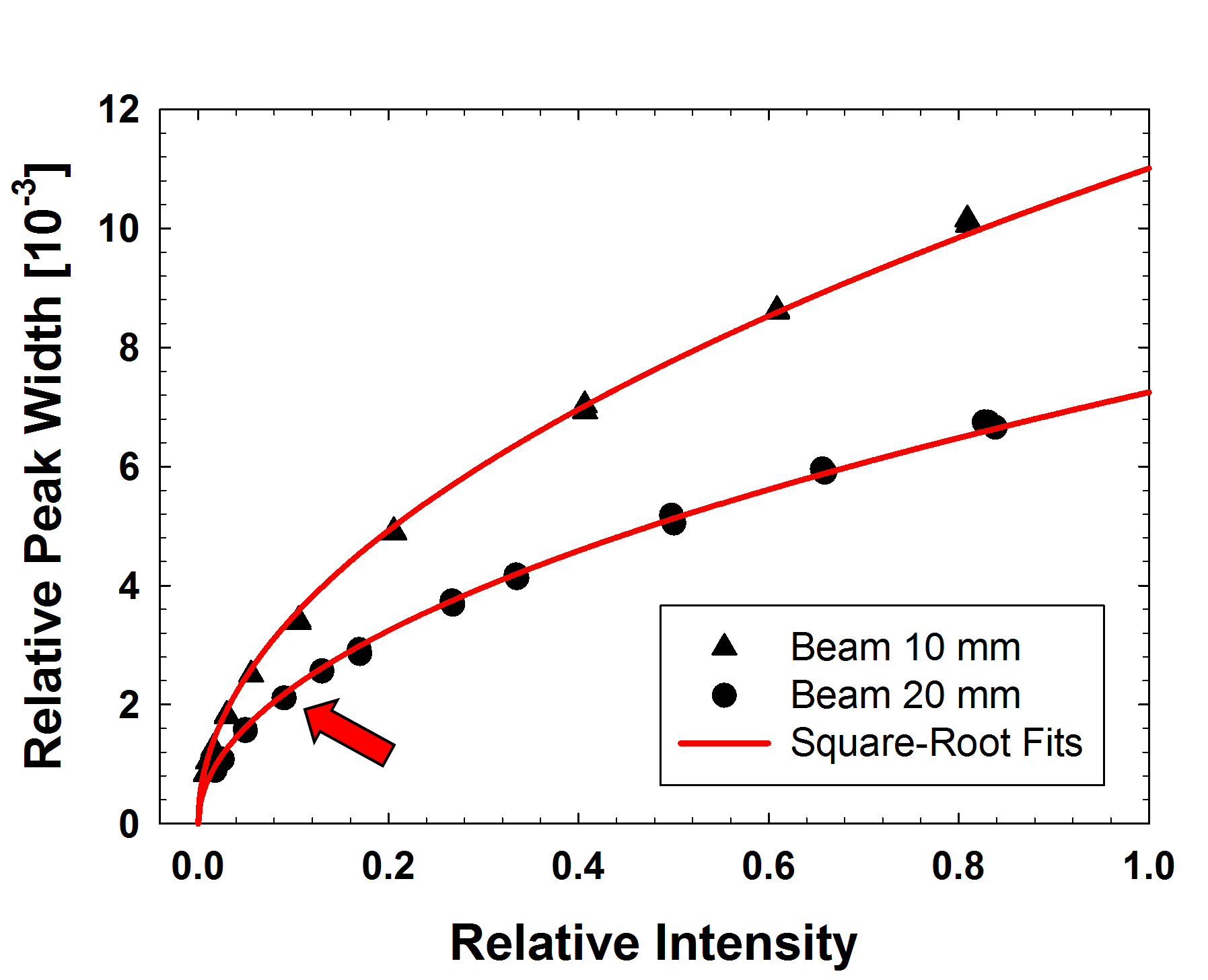}
	\caption{Relative width of the histogram peaks $\delta I / I_0$ for both aperture settings. The exposure times of the employed images range from 50~ms up to 8~s (10~mm aperture) and up to 5~s (20~mm aperture), respectively. The lines represent fits to the data points using the square-root fit function. As an example, the shown arrow indicates the data point deduced from the plot presented in Fig.\ \ref{fig:Fig02_Histograms20mm} bottom.} 
	\label{fig:Fig03_PeakWidth}
\end{figure}\\
In Fig.\ \ref{fig:Fig03_PeakWidth} the relative peak width $\delta I / I_0$, i.e.\ the standard deviation of the Gaussian fits to the intensity histograms, as a function of the relative intensity $I / I_0$ is presented for both neutron beam aperture settings.
The data is fitted most generally using the power function: $f(x) = a \cdot x^p + b$. For a pure Poisson-like behavior, it is expected that $p = 0.5$ and the offset $b=0$.
Performing a least squares fit, results in offsets in agreement with zero within the error of the fit and $p = (0.53 \pm 0.01)$ for both aperture settings. The latter value deviates slightly from 0.5, however, this can be explained due to the residual statistical noise of the averaged normalization image(s). A systematic study of the power parameter shows that $p$ converges to 0.5 for an increasing number of images employed to obtain the normalization. 
In the further evaluation, the power function is simplified to the square-root fit function with only one free parameter: $f_{\text{sq}} (x) = a \cdot \sqrt{x}$.
Hence, together with Eq.\ (\ref{eq:relativeintensity}) it follows:
\begin{equation}
	\alpha = a^2 \cdot t_{\text{N}} \cdot \dot{I}_{d}
	\label{eq:alpha-result}
\end{equation}
where $\dot{I}_{d}$ is the mean accumulated intensity per exposure time for the corresponding aperture diameter $d$ and $I_0 = t_{\text{N}} \cdot \dot{I}_{d}$. 
The results of the fits are summarized in Table\ \ref{tbl:fitresults}, yielding a value for $\alpha$ of about 1.68. 
Since $\alpha$ represents a factor solely describing the detection system, it is independent of the size of the neutron beam aperture. Employing Eq.\ (\ref{eq:intensity-error2}) now allows to directly provide a corresponding statistical error for each pixel of a radiographic image with a given intensity $I$.\\
From the same analysis the detected neutron flux per SINQ proton beam current at the scintillator screen can be extracted using Eq.\ (\ref{eq:intensity}) and (\ref{eq:alpha-result}):
\begin{equation}
	\Phi_{\text{det}} = \frac{ \epsilon \dot{N}}{A_{\text{eff}}  \cdot I_{\text{SINQ}}} = \frac{1}{a^2 \cdot t_{\text{N}} \cdot A_{\text{eff}} \cdot I_{\text{SINQ}}}
	\label{eq:detectedneutronflux}
\end{equation}
The corresponding results are also presented in the table.
The values can be compared to gold foil activation analysis performed at the camera position in the year 2006 \cite{Kaestner2011387}.
Both results are in agreement within a small factor. However, even if one assumes a perfect detection efficiency of $\epsilon=1$, the new values exceed the previous gold foil measurements by a factor $1.6-1.7$. This increase can probably be fully explained by the improved neutron flux due to upgrades of the SINQ spallation target and additionally by a different susceptibility to the neutron energy spectrum of the gold foil and the scintillator.
\begin{table*}
	\centering
		\begin{tabular}{ >{\centering\arraybackslash} p{1.25cm} >{\centering\arraybackslash} p{1.5cm} >{\centering\arraybackslash} p{1.25cm} >{\centering\arraybackslash} p{1.75cm} >{\centering\arraybackslash} p{1.75cm} >{\centering\arraybackslash} p{3cm} >{\centering\arraybackslash} p{3cm}}
			\hline
			$d$~[mm]   &   $\dot{I}_d$~[s$^{-1}$]   &   $t_{\text{N}}$~[s]   &  $a$~$[10^{-3}]$  &   $\alpha$   &   $\Phi_{\text{det}}$~[cm$^{-2}$s$^{-1}$mA$^{-1}$] & $\Phi_{\text{Au}}$~[cm$^{-2}$s$^{-1}$mA$^{-1}$] \\
			\hline       
			  10    & $1381 \pm 1$  & 10 & $11.0 \pm 0.1$  & $1.67 \pm 0.03$ & $(5.5  \pm 0.1) \times 10^6$ & $3.2 \times 10^6$ \\
				20    & $5321 \pm 11$ & 6  & $7.25 \pm 0.05$ & $1.68 \pm 0.02$ & $(21.1 \pm 0.3) \times 10^6$ & $13  \times 10^6$ \\
			\hline
		\end{tabular}
	  \caption{Overview of the fit results and deduced values for $\alpha$ and $\Phi_{\text{det}}$ for both aperture settings. As a comparison also the flux determined by gold foil activation measurements $\Phi_{\text{Au}}$ is presented \cite{Kaestner2011387}.}
	  \label{tbl:fitresults}
\end{table*}



\section{Conclusion}

A new fast and convenient calibration procedure for neutron detection systems has been presented. It can be used to determine the statistical uncertainty of the intensity due to neutron counting statistics for each individual pixel of a radiographic image. The technique was applied to data recorded at the cold neutron imaging beam line ICON at the Paul Scherrer Institute. From the data the expected characteristic Poisson-like behavior of the statistical uncertainty was obtained.
In addition, from the analysis an effective neutron flux was determined, which is comparable with results previously obtained from gold foil activation measurements.
The presented method will prove invaluable for systematic studies concerning signal-to-noise optimization and for quantitative image analysis, e.g.\ regression analysis.\\

The present work is supported by the Swiss National Science Foundation under grant number $200020 \_ 159754$.



\end{document}